\PassOptionsToPackage{round,sort&compress}{natbib}
\PassOptionsToPackage{table}{xcolor}

\documentclass{iopjournal}

\renewcommand{\articletype}[1]{{\vspace*{-8mm}\noindent \Large \sf \phantom{Journal Name}}

\vspace*{8mm}\noindent\reversemarginpar{\scriptsize \sf{\bfseries \MakeUppercase{#1}}}}

\usepackage{graphicx}
\usepackage{booktabs}
\usepackage{tabularx}
\usepackage{array}
\usepackage{enumitem}
\usepackage{xcolor}
\usepackage{listings}
\usepackage{natbib}
\usepackage{xurl}                          
\usepackage{hyperref}
\usepackage{doi}                           

\definecolor{Teal}{HTML}{0F7C7C}
\definecolor{TealDark}{HTML}{0B5E5E}
\definecolor{Slate}{HTML}{666F78}
\definecolor{Hairline}{HTML}{D7DADD}
\definecolor{CodeBg}{HTML}{F6F7F8}
\hypersetup{
  colorlinks=true, linkcolor=teal, citecolor=teal, urlcolor=teal,
  pdftitle={WFDB-x: Native Implementations of the WFDB Software Package in Seven Programming Languages with Unit-Tested Behavioral Equivalence},
  pdfauthor={Reza Sameni},
  pdfkeywords={WFDB, PhysioNet, physiological signals, ECG, open-source software, conformance testing}
}

\lstdefinestyle{code}{
  basicstyle=\ttfamily\footnotesize,
  columns=fullflexible,
  keepspaces=true,
  showstringspaces=false,
  breaklines=true,
  breakatwhitespace=true,
  postbreak=\mbox{\textcolor{Slate}{$\hookrightarrow$}\space},
  backgroundcolor=\color{CodeBg},
  frame=single,
  rulecolor=\color{Hairline},
  framesep=6pt,
  xleftmargin=4pt,
  xrightmargin=4pt,
  commentstyle=\color{Slate}\itshape,
  keywordstyle=\bfseries,
  stringstyle=\color{TealDark},
  upquote=false,
  aboveskip=8pt,
  belowskip=8pt
}
\newcommand{\wfdbx}{WFDB-x}
\newcommand{\code}[1]{\texttt{#1}}

\begin{document}

\articletype{}

\title{\wfdbx{}: Conformance-Tested Implementations of the WFDB Software Package in Seven Programming Languages}

\author{Reza Sameni\textsuperscript{1,2}\orcid{0000-0003-4913-6825}}

\affil{\textsuperscript{1}Department of Biomedical Informatics, Emory University, Atlanta, GA, USA}

\affil{\textsuperscript{2}Department of Biomedical Engineering, Georgia Institute of Technology and Emory University, Atlanta, GA, USA}

\email{rsameni@emory.edu}


\begin{abstract}
\textit{Objective:} The WFDB format is widely used to store and exchange physiological waveforms and annotations, including many PhysioNet datasets. However, existing software implementations are limited to C, Python, and MATLAB/Octave (no longer maintained), can produce different results in corner cases and checksum calculations, and have defaults such as sampling frequency and gain, that can lead to silent misinterpretation of recordings. We developed \wfdbx{}, a cross-language, performance-optimized reimplementation of WFDB in seven popular programming languages, with extensive unit tests, to provide consistent WFDB processing across programming languages.

\textit{Approach:} We reimplemented the WFDB Software Package 10.7.0 natively in Python, C++, Rust, Julia, R, MATLAB/Octave, and Kotlin, using AI-assisted development under human direction and review. Each implementation provides library functions, a high-level interface, and command-line tools. We compared their outputs with those of the original C programs using 1098 conformance cases, 877 legacy-mode cases, unit tests, and differential fuzzing. Deliberate changes remove implicit waveform parameters, restrict file and network access, and prevent annotation loss. A legacy mode retains the original behavior.

\textit{Main results:} All seven implementations passed every conformance and legacy-mode case on the tested Linux platform, with matching outputs. The unit tests contain 9353 assertions. C++, Rust, and Julia read records 1.5--6.9 times faster than the original C library, while the Python implementation wrote records 9.2 times faster than the existing wfdb-python library in the benchmark.

\textit{Significance:} \wfdbx{} provides a tested, open-source basis for reading and writing physiological recordings consistently across seven languages, supporting reproducible research and software development without requiring a shared programming environment. We also share details of the AI-assisted development process and token utilization, which may be useful for other applications.
\end{abstract}


\section{Introduction}
Advances in physiological signal processing depend on the availability of well-annotated, publicly accessible data. A cardiac beat detector, an arrhythmia classifier, or a sleep-staging model is only as credible as the recordings and reference annotations used to develop and evaluate it. Reproducible research and reliable comparisons between methods also require standardized data formats and software that consistently read, process, and write physiological signals and annotations. Most modern data formats also insert some form of checksum to check data integrity. For more than four decades, the WFDB (WaveForm DataBase) format and its associated software package have served this purpose. Originally developed by George B. Moody and team at the Massachusetts Institute of Technology (MIT), the WFDB Software Package provides tools for storing, accessing, and processing physiological recordings and their annotations~\citep{moody2022wfdb}. The earliest functions of its C library date from 1983, and its latest release, version 10.7.0, was published in 2022.

The WFDB format originated with the MIT-BIH Arrhythmia Database, recorded between 1975 and 1979 and distributed beginning in 1980. The database contains 48 half-hour excerpts of two-channel ambulatory electrocardiograms (ECGs), with approximately 110,000 beat annotations, and became one of the first widely available benchmarks for arrhythmia detection algorithms~\citep{moody2001mitbih}. PhysioNet, established in 1999, made this and many subsequent databases freely available, together with the WFDB software needed to access them~\citep{goldberger2000physionet,moody2001physionet}. As of 2026, PhysioNet hosts 457 databases and has registered users from more than 180 countries~\citep{mitnews2026physionet,pollard2026physionet}. The data resources include the MIT-BIH databases, waveform recordings associated with the MIMIC critical-care databases~\citep{johnson2016mimic3,johnson2023mimic4}, and the 21,837 twelve-lead ECGs of PTB-XL~\citep{wagner2020ptbxl}. These datasets are distributed as WFDB records, consisting of a text header (\code{.hea}) describing the signals, one or more binary signal files (\code{.dat}), and annotation files identifying beats, rhythms, and other events.

The WFDB format also supports the George B. Moody PhysioNet Challenges, held annually since 2000 and attracting hundreds of participating teams worldwide~\citep{moodychallenge,clifford2026physionet}. Many of these Challenges distribute their data in WFDB format, and their datasets, reference annotations, and submitted algorithms have become widely used benchmarks. The separation of data, metadata, and annotations has been one of the key features of WFDB, providing flexibility in linking different modalities, updating records efficiently, and deidentifying data in data challenges. Consequently, a substantial body of physiological signal processing research relies on software implementations of WFDB to access, exchange, and analyze data.

Despite the widespread use of WFDB, consistent handling of its records across programming languages is not guaranteed. The original software package is written in C, while researchers increasingly work in Python, MATLAB, R, Julia, and other environments. Existing alternatives either reimplement the format independently or rely on wrappers around the original C programs. Differences in their handling of signal formats, numerical conversions, annotations, and errors can produce inconsistent results. Moreover, some historical design choices of the C package, including implicit assumptions about sampling frequency, signal gain, resolution, and physical units, can silently affect the interpretation of recordings. These differences matter because the results of an analysis depend not only on the algorithm but also on the samples, scaling factors, timing, and annotations returned by the software. Errors or inconsistencies introduced when reading a record can propagate through an entire processing pipeline without being apparent in the final results. As a result, many of the public datasets in WFDB format include inconsistent checksums, irrelevant scaling factors and inaccurate sampling frequencies in their headers.

To address these issues, we developed \wfdbx{}, a cross-language AI-assisted reimplementation of the WFDB Software Package, version 10.7.0, implemented natively (not as caller wrappers) in seven programming languages: Python, C++, Rust, Julia, R, MATLAB/Octave, and Kotlin~\citep{wfdbx2026}. Unlike independent implementations of the format, \wfdbx{} uses the original C source code as its reference and reproduces its behavior through a common conformance-testing framework. All seven implementations provide the most commonly used features of the WFDB library, including the 33 command-line tools for reading, writing and converting signals and annotations, and a high-level programming interface, without depending on the C library or on another programming language. The outputs of every implementation are verified against those of the reference C implementation, including byte-for-byte comparisons of the files it writes, with explicitly documented exceptions. \wfdbx{} also corrects a small set of legacy behaviors that can cause silent errors, unsafe file access or loss of annotations, and keeps the original behavior available through a legacy compatibility mode. The analysis, plotting and visualization programs of the C package, such as its beat detectors and evaluation tools, are outside the scope of \wfdbx{}. A broader collection of these tools are available in our open-source electrophysiological toolbox (OSET)~\citep{oset}. The result is a consistent implementation of WFDB in seven languages, so that physiological signal processing can be reproduced without requiring researchers to use the same programming environment or include extra packages in deployed software. Each implementation has also been optimized for speed without changing its results: the C++, Rust and Julia implementations read records 1.5 to 6.9 times faster than the original C library, and the Python implementation writes records 9 times faster than wfdb-python (Section~\ref{sec:performance}).

Section~\ref{sec:background} reviews existing WFDB implementations and the limitations that motivated this work. Section~\ref{sec:goals} presents the design objectives, and Section~\ref{sec:architecture} describes the software architecture. Section~\ref{sec:divergences} explains the deliberate departures from the original C package. Sections~\ref{sec:development} and \ref{sec:verification} describe the development process and verification framework, respectively. Section~\ref{sec:performance} evaluates performance, and Section~\ref{sec:discussion} discusses the methodology and limitations. The appendices provide examples of reading and writing signals and annotations in each language and using the command-line tools.

\section{Background and motivation}
\label{sec:background}

\subsection{The WFDB Software Package}

The WFDB Software Package~\citep{moody2022wfdb} includes a C library and more than 70 applications. The library reads and writes headers, signal files in about a dozen storage formats (from 8-bit first differences to the bit-packed 12-bit format 212 of the MIT-BIH recordings and, since 2020, FLAC-compressed formats), multi-segment and multi-frequency records, annotation files in the MIT and AHA formats, and EDF files~\citep{kemp1992edf,kemp2003edfplus}. It has tools that read and write samples and annotations (\code{rdsamp}, \code{wrsamp}, \code{rdann}, \code{wrann}), resample and convert records (\code{xform}, \code{snip}), and convert between WFDB and other formats (\code{edf2mit}, \code{mit2edf}, \code{wav2mit}, \code{wfdb2mat} and others). The C package serves as the reference implementation. Its behavior is described in the WFDB Programmer's Guide and the WFDB Applications Guide~\citep{moody2022wfdb}.

Although the C library is portable, using it from another programming language requires platform-specific builds and language bindings. Much of today's physiological signal processing and machine learning is performed in Python, MATLAB, Julia, and R (particularly in biostatistics communities), while mobile applications have additional deployment requirements in languages such as Kotlin and Java. Rust has also gained popularity for systems programming and performance-critical applications due to its memory safety and efficiency.

\subsection{wfdb-python and the WFDB Toolbox for MATLAB and Octave}

Two alternatives to the C library are widely used. \emph{wfdb-python}~\citep{xie2023wfdbpython}, developed at the MIT Laboratory for Computational Physiology, is a native Python library. Rather than translating the C code, it implements the format independently and provides its own interface (\code{rdrecord}, \code{rdsamp}, \code{rdann}, \code{wrsamp}). The \emph{WFDB Toolbox for MATLAB and Octave}~\citep{silva2014wfdbmatlab}, created in 2012, provides MATLAB and Octave access to the WFDB applications through MATLAB and Java wrappers that call compiled C libraries, rather than implementing the WFDB formats natively. Its latest published version, 0.10.0, was last updated in 2017 and published on PhysioNet in 2021~\citep{wfdbmatlab0100}. The toolbox is no longer actively maintained~\citep{wfdbio_matlab}.

These independent implementations improve accessibility but differ from the C library in certain behaviors. One important example concerns checksums. A WFDB header can contain a 16-bit checksum of the samples of each signal. The C library computes it as the sum of the integer samples, allowed to overflow, and writes it as a signed 16-bit value; when the samples read do not match this checksum, it reports ``\code{checksum error in signal}''. The existing implementations do not fully reproduce this 16-bit overflow design: each reduces the sum to 16 bits in its own way, so that the same samples can receive different checksums depending on the software that wrote the header. Moreover, in our test, wfdb-python 4.3.1 read a record whose samples had been modified after its header was written without issuing a warning, and wfdb-python's issue tracker reports a conversion function that computes checksums from physical values rather than from the stored samples~\citep{wfdbpython_issue593}. A reader that does not verify the checksum cannot alert the user to a mismatch. As a result, the software used to read or write a record may affect subsequent analyses without any visible indication. Many of the publicly available WFDB datasets created by wfdb-python are affected by this issue.

These differences make it difficult to determine whether reading or writing the same record in different programming environments produces the same samples, annotations, and files as the original C tools.

\subsection{Legacy design choices and their consequences}
\label{sec:aged}

Several defaults and file-handling conventions in the C package reflect the databases and computing environments of the 1980s. In modern workflows, they can lead to silent errors that are difficult to detect.

\subsubsection{Built-in signal parameters:} When a record does not give a sampling frequency, the C library assumes 250\,Hz (\code{WFDB\_DEFFREQ}), the frequency of the AHA database. A missing gain becomes 200 analog-to-digital units (adu) per millivolt, a missing ADC resolution 12 bits, and missing units millivolts. The WAV and AHA converters have their own built-in frequencies and gains. These defaults reflect the characteristics of early databases. Modern recordings are sampled at anything from a few hertz to tens of kilohertz, with 16- or 24-bit converters, and are often not ECGs. When a defaulted parameter differs from the actual value, the resulting time intervals, resampled signals, or physical amplitudes are incorrect, with no indication that a default was used. 

\subsubsection{The WFDB path:} The C library finds a file by searching a list of locations, the WFDB path, whose default ends with PhysioNet's server. For example, \code{rdsamp -r 100} reads record 100 from whichever location has it first and downloads it if no local copy exists. Although convenient for interacting with PhysioNet datasets, this search procedure can select an outdated local copy, initiate an unexpected network request, or access different records when the same script is run on different machines. The library also asks the server for an optional \code{.info} file of every record it opens; for an absolute record name, the request contains the full local path. A crafted header can name signal files elsewhere on the file system. Finally, the network entry of the default path, \code{http://physionet.org/physiobank/database}, dates from PhysioNet's former PhysioBank layout; PhysioNet now serves each database from a versioned location, such as \code{https://physionet.org/files/mitdb/1.0.0/} for the MIT-BIH Arrhythmia Database, so that the default path no longer names the files of a specific database version.

\subsubsection{Annotation sorting:} The C library delegates the sorting of out-of-order annotation files to a separate program, \code{sortann}, which it runs through the system shell when the file is closed. A library function therefore depends on the environment in which it runs: on whether, and which, \code{sortann} is installed, and on the shell's interpretation of the record names it is given. Moreover, \code{sortann} can silently discard annotations. When an annotation at the very start of a record (time 0, annotation number 0, channel 0) arrives after later annotations, it is lost together with the annotations placed behind it; for example, annotations written at times 5, 0 and 3 produce a file that contains only the first. Since annotations are the reference labels against which detectors and classifiers are evaluated, such silent losses directly affect the validity of downstream results.

\subsubsection{Malformed input and write errors:} Some malformed headers and EDF files cause the C programs to crash, attempt to allocate gigabytes, or give results that depend on the machine. Tools that fail to write their output, for example on a full disk, report the error and still exit with status 0, so that a script cannot detect the failure.

Many of these choices served the original applications of WFDB and helped sustain its use for decades. The relatively minor bugs are also a testament to the robustness of software developed and maintained over four decades by George B. Moody and team. Nevertheless, a new implementation should address their consequences in current workflows while retaining backward compatibility with the original software.

\section{Design objectives}
\label{sec:goals}

We set the following design requirements before implementing \wfdbx{}:

\begin{enumerate}
  \item \textit{Behavioral equivalence to the original C code:} Each implementation must read and write the same files as the C library. Command-line tools must reproduce the outputs, messages, exit statuses, and generated files of their C counterparts, except for documented deviations. The C source code, rather than its documentation or another implementation, serves as the reference.
  \item \textit{Consistency across languages:} The implementations must produce identical results for the same inputs, regardless of programming language.
  \item \textit{Completeness:} Each implementation must provide the full WFDB library and all signal, annotation, and conversion tools. Interactive viewers (WAVE), plotting tools, and the analysis programs in the package's \code{psd} directory, that are readily covered by existing tools such as OSET~\citep{oset}, and the PhysioNet Cardiovascular Toolbox~\citep{Vest2018}, are outside the scope of this work.
  \item \textit{Native implementation with minimal dependencies:} Each implementation must be written in its own language, without calling the C library or another language. Dependencies are limited to the language's standard library, except for NumPy~\citep{harris2020numpy} in Python. However, because performance varies across languages, we also support cross-language function calls and provide guidelines for developing language wrappers.
  \item \textit{Efficiency:} Codebase optimization and performance improvements must not change the results.
 \item \textit{Documentation and testing:} Each implementation must include tests and installation instructions. The Python implementation must also be installable from the Python Package Index (PyPI). We also plan to distribute the R implementation through the Comprehensive R Archive Network (CRAN)~\citep{cran,hornik2012cran} and the Julia implementation through the Julia General registry~\citep{julia_general,bezanson2017julia}.
  \item \textit{No default or implicit waveform parameters or physical units:} Where the C implementation assumes default values for unspecified waveform parameters or physical units (Section~\ref{sec:aged}), \wfdbx{} must remove these assumptions, document the changes and their rationale, and provide a legacy mode to reproduce the original behavior when explicitly requested by the user.
\end{enumerate}

Strict compatibility also requires reproducing certain behaviors of the C package that may be unintended, such as platform-dependent \code{printf} formatting and the reuse of previously read values when input is incomplete. These behaviors are documented in the repository, separately from the deliberate changes made in \wfdbx{}.

\section{Software architecture}
\label{sec:architecture}

\begin{figure}[t]
  \hspace*{-20mm}%
  \begin{minipage}{\dimexpr\textwidth+20mm\relax}
    \includegraphics[trim={0 2cm 0 0}, clip,width=\linewidth]{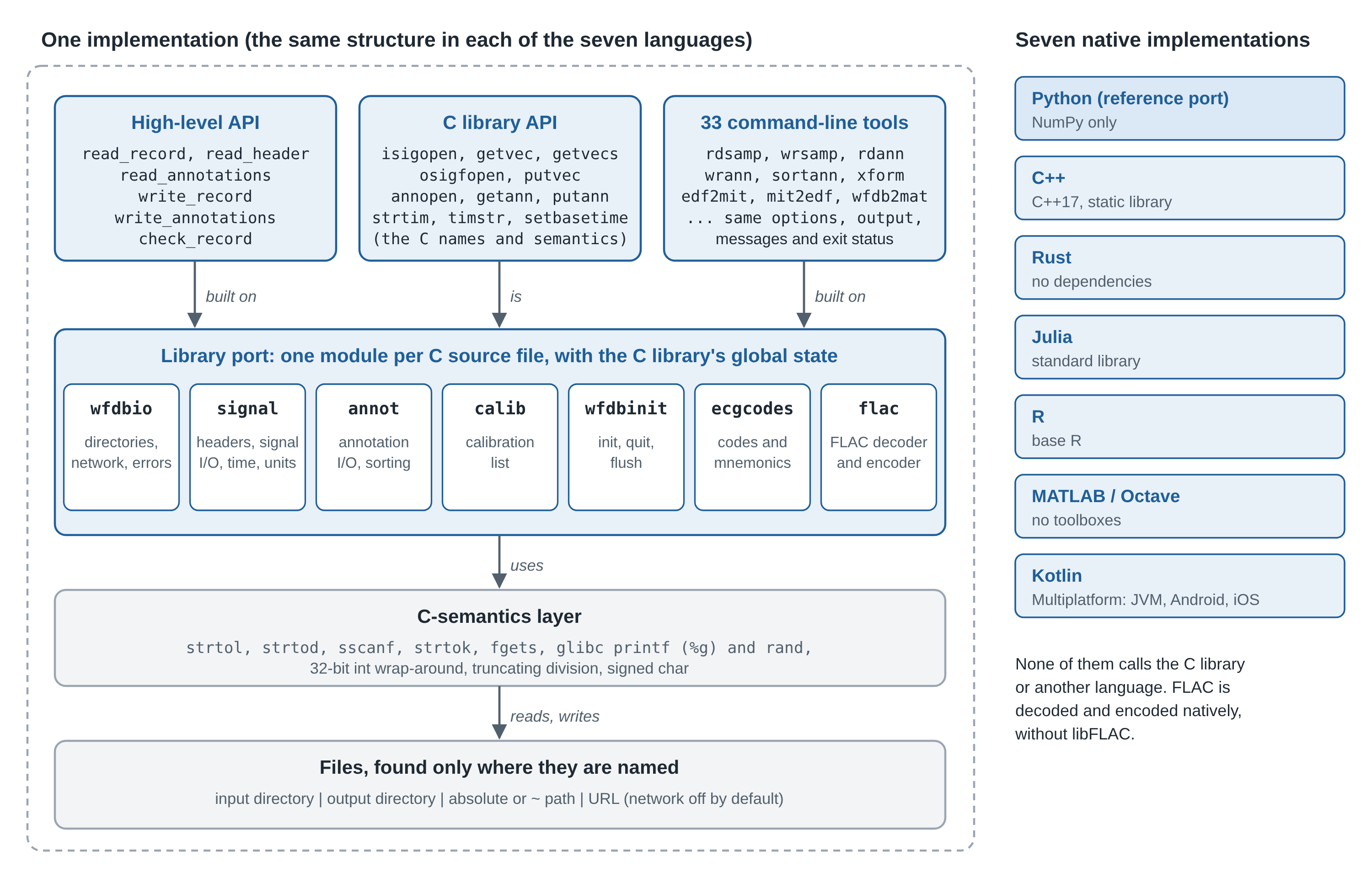}
    \caption{The structure of one \wfdbx{} implementation. The high-level API and the 33 command-line tools are built on a native reimplementation of the WFDB C library, written in the language of the implementation with one module per C source file, which keeps the C library's function names and global state. Beneath it, a layer reproduces the C semantics on which the output depends. All seven implementations have this structure.}
    \label{fig:architecture}
  \end{minipage}
\end{figure}


\subsection{Programming interfaces}
Each language implementation has three interfaces (Figure~\ref{fig:architecture}), all written natively in the language of the implementation; none of them calls the original C library or other languages.
\begin{enumerate}
  \item \textit{A high-level API}: \code{read\_record}, \code{read\_header}, \code{read\_annotations}, \code{write\_record}, \code{write\_annotations} and \code{check\_record} (\code{readRecord} and so on in Kotlin). A single function call reads a whole record into an array (a NumPy array, a Julia matrix, an R matrix, etc.) or writes one, with named options and errors reported as exceptions or error values. The function \code{check\_record} reports problems in a record, such as samples out of range, checksums that do not match and inconsistent lengths, similar to the \code{wfdbcheck} toolbox~\citep{wfdbcheck}. This layer has no counterpart in the original C library.
  
  \item \textit{The WFDB library functions}: a reimplementation, in each language, of the functions of the C library, under their C names and with their C behavior (\code{isigopen}, \code{getvec}, \code{getann}, \code{putann}, \code{strtim}, \ldots). For example, \code{getvec} returns one frame of samples per call in Python, C++, Rust, Julia, R, MATLAB/Octave and Kotlin, exactly as it does in C. Arguments follow the conventions of each language: in R and MATLAB/Octave, for instance, values that the C function returns through pointer arguments are returned as results. This interface serves to port programs written for the C library and to process a record as a stream, one frame or one annotation at a time.
  
  \item \textit{The 33 command-line tools}: 18 signal and annotation tools (\code{rdsamp}, \code{wrsamp}, \code{rdann}, \code{wrann}, \code{sortann}, \code{xform}, \code{snip} and others) and 15 format converters (\code{edf2mit}, \code{mit2edf}, \code{wav2mit}, \code{mit2wav}, \code{wfdb2mat}, \code{parsescp} and others), with the options, output, messages and exit status of the C tools. Appendix~\ref{app:tools} lists them.
\end{enumerate}

The high-level API and the command-line tools are both built on the library functions of the second interface. The high-level API therefore returns the same samples and annotations as these functions and the original C library. For a more robust programming experience and to mitigate the risks and shortcomings of the original implementation, new applications are encouraged to use the high-level API.

\subsection{A file by file translation of the C library}
The library of each implementation is a translation of the WFDB C library, organized in a similar way. Each source file of the C library corresponds to one module in each language: \code{lib/signal.c}, for example, corresponds to \code{\_signal.py} in Python, \code{signal.cpp} in C++, \code{signal.rs} in Rust, \code{signal.jl} in Julia, \code{signal.R} in R and \code{Signal.kt} in Kotlin. MATLAB/Octave, which requires one file per public function, has instead one file per C function (\code{isigopen.m}, \code{getvec.m}, \ldots). Each translated function has the name of its C counterpart and serves the same purpose. The library also keeps the same state between calls as the C library, such as the open signal and annotation files and the current sampling frequency. Each C application, such as \code{app/rdsamp.c}, is translated into a command-line tool of the same name, which calls these functions in the same order as the C program calls the C library.

This one-to-one correspondence makes each function and each tool directly comparable with its C source. As a result, in future development and validation, when an output differs from that of the C package, the difference can be traced to a specific function/file. The header comment of every source file names the C file it translates, its original authors and copyright, and states how, why and where in the code it differs from the C implementation.

\subsection{Reproducing C semantics}
Matching the C programs byte for byte requires more than implementing the same algorithms. It also requires the same integer arithmetic, parsing of arguments (header files, command-line options, and text input/output), and text formatting as C on the reference platform (x86-64 Linux with glibc, in our tests). Each implementation except C++ therefore contains a small internal compatibility layer that reproduces these C behaviors: 32-bit integer wrap-around, which determines checksums and the samples read from format 8, a formerly popular format that stores each sample as an 8-bit difference from the previous one (allowing more efficient encoding of waveforms with a large dynamic range); truncating integer division; signed \code{char}; the parsing functions \code{strtol}, \code{strtod}, \code{sscanf}, and \code{strtok}, including the values that \code{sscanf} leaves unchanged when its input ends early, on which several tools depend; \code{fgets} with its fixed buffer sizes; glibc's correctly rounded \code{printf} formatting for \code{\%g}, \code{\%f}, and \code{\%e}; and glibc's \code{rand}, which generates the dither used by \code{wrsamp} and \code{xform}. The C++ implementation calls the C standard library directly for parsing and formatting and reproduces only glibc's \code{rand} and \code{strtok}, whose behavior differs between C libraries, so that its results do not depend on the platform.

Finally, all implementations treat text as a sequence of bytes, as C does: files and standard input and output are read and written in binary mode, without decoding into Unicode and without converting line endings, so that every byte, including non-ASCII characters in signal descriptions and annotation text and the \code{\textbackslash r\textbackslash n} line endings of header files, is preserved exactly.

\subsection{FLAC without libFLAC}
Since version 10.7.0 (2022), the WFDB C package can store signals compressed with FLAC, a lossless compression format (signal formats 508, 516 and 524)~\citep{vanBeurden2024}. To do so, it depends on an external library, libFLAC, which must be installed before the package is compiled. \wfdbx{} does not have this dependency: each implementation includes its own FLAC decoder, translated from that of libFLAC 1.4.3, and its own FLAC encoder. FLAC defines how a file is decoded, but leaves the encoder free to choose how to compress the samples, for example the block size and the prediction method. Two encoders can therefore produce different files from the same samples, both of which decode to exactly those samples. For this reason, the tests compare the FLAC files that \wfdbx{} writes with those of the C package by their decoded samples.

\subsection{Optimized block reading}
The WFDB C library reads a record one frame at a time: each call of \code{getvec} returns one sample of each signal. Reading a long record in this way costs one function call and one small copy per frame, which is significant in every language and even more significant in interpreter languages (Python, R and MATLAB/Octave). \wfdbx{}'s \code{read\_record} therefore reads signal files through a block reader (\code{getvecs} in C++, Rust, Julia and Kotlin; \code{getframes} in Python, R and MATLAB/Octave), which returns many frames per call. It decodes the stored samples a block of frames at a time, and then processes them signal by signal, in tiles of 1024 frames that stay in the processor's cache (C++, Rust, Julia and Kotlin) or with whole-array operations (Python, R and MATLAB/Octave).

Multi-frequency records (such as data collected from ICU bedside monitors with different sampling frequencies per modality) need one more step. In such a record, some signals have several samples in each frame (for example, an ECG sampled at 500\,Hz in a record whose frames are at 125\,Hz has 4 samples per frame). By default, \code{getvec} returns a single value per signal and frame: for a signal with several samples in the frame, the mean of these samples, rounded to the nearest integer (halves rounded up), or the invalid-sample value if any of them is invalid. Every signal is therefore returned at the frame rate. The block reader computes the same averages for all the frames of a block; signals with one sample per frame are copied unchanged. The block reader returns exactly what the same number of \code{getvec} calls would return, including the checksum verification, the handling of the end of the record and the rest of the library's state. When physical units are requested, \code{read\_record} converts the samples after reading them, without intermediate copies of the data.

In each language, randomized equivalence tests check every fast path against the frame-by-frame code it replaces, and the conformance suite unit-test checks every tool against the C programs.

\subsection{Choice of languages}
The seven languages were chosen to cover the major programming languages most widely used in healthcare applications~\citep{Seyedhasani2025}, including those used for physiological signal analysis, and to include both compiled and interactive languages:

\begin{itemize}
  \item \textbf{Python} is widely used in data science and machine learning. In \wfdbx{}, the Python implementation was written first and serves as the reference for the cross-language comparisons; the other implementations follow its structure. \wfdbx{}'s Python implementation represents samples as NumPy arrays~\citep{harris2020numpy}, its only dependency, and is distributed on PyPI as \code{wfdb-x}~\citep{Sameni2026WFDBX}.

  \item \textbf{C++} (C++17)~\citep{iso2017cpp} and \textbf{Rust}~\citep{matsakis2014rust} are compiled languages, used for systems software and for applications where speed and memory use matter. Like the other implementations, \wfdbx{} provides the complete library, the high-level API and the command-line tools, and has no dependency beyond the language's standard library. Compiled to machine code, they are among the fastest of the seven implementations for reading and writing and use the least memory (Section~\ref{sec:performance}), and their command-line tools start almost as fast as the WFDB C tools (about 2\,ms, against 1.3\,ms). Because the C++ implementation keeps the function names of the C library, programs written for the C library can be ported to it with few changes. Rust guarantees memory safety at compile time, without a garbage collector~\citep{jung2018rustbelt}, which prevents classes of errors that C and C++ code can contain, such as buffer overflows and accesses to freed memory, while keeping a performance comparable to that of C and C++.

  \item \textbf{Julia}~\citep{bezanson2017julia} is a language for scientific and numerical computing, designed to combine the interactive use of Python, R or MATLAB with the speed of compiled code. It compiles each function at its first call, so it suits numerical work in long-running sessions, where the compiled code is reused.

  \item \textbf{R}~\citep{rcore2026} is widely used for statistics and biostatistics, including clinical and epidemiological research. \wfdbx{} implements it in base R, without additional packages, and returns signals as R matrices, ready for R's statistical functions.

  \item \textbf{MATLAB/Octave}: MATLAB is widely used for signal processing in engineering research and teaching, and GNU Octave is a free language largely compatible with it. \wfdbx{} implements the subset of the language common to the two, without relying on any MATLAB toolboxes. It replaces the WFDB Toolbox for MATLAB and Octave~\citep{silva2014wfdbmatlab}, which is no longer maintained, with a native implementation.

  \item \textbf{Kotlin Multiplatform}: Kotlin is the language Google recommends for Android development~\citep{android_kotlin}, and Kotlin Multiplatform compiles the same code to the Java Virtual Machine (JVM), including Android, and to native code for iOS. Many wearable and mobile health devices export their recordings either as text files, such as comma-separated values (CSV), which are large and slow to read, or in proprietary binary formats that other vendors' software cannot read. \wfdbx{}'s Kotlin implementation addresses this need; it enables mobile and embedded applications, including those of wearable and bedside devices, to read and write recordings directly in the standard WFDB format, without a native C library to build for each platform.
  
  \end{itemize}

\section{Design decisions and deliberate departures from the original WFDB C package}
\label{sec:divergences}

\wfdbx{} follows the original C package except for the deliberate changes summarized in Table~\ref{tab:divergences}. These changes address the problems described in Section~\ref{sec:aged} and are implemented consistently in all seven languages. The author and collaborators selected them after examining the C behavior, its original purpose, and possible alternatives.

The repository includes the unmodified WFDB Software Package 10.7.0 sources in the \code{third\_party} directory, under their original licenses: the GNU Library General Public License, version 2 or later, for the library, and the GNU General Public License, version 2 or later, for the applications. A patch in \code{tests/reference} applies the deliberate changes to a copy of these sources. The patched C programs provide the expected outputs for the cross-language conformance suite; they are not intended as a maintained or recommended WFDB distribution. The only exceptions are 20 test cases in which the C programs crash or produce machine-dependent results. For these cases, \wfdbx{} defines a consistent result (Section~\ref{sec:verification}). Our deliberate design choices and divergences are detailed below.

\begin{table}[tb]
\centering
\small
\caption{Deliberate departures of \wfdbx{} from the WFDB Software Package 10.7.0. Legacy mode restores the original behavior in each case.}
\label{tab:divergences}
\begin{tabularx}{\textwidth}{>{\raggedright\arraybackslash}p{2.8cm}>{\raggedright\arraybackslash}X>{\raggedright\arraybackslash}X}
\toprule
Topic & C package & \wfdbx{} \\
\midrule
Signal parameters & Assumes 250\,Hz, 200~adu/mV, 12 bits and mV when a record or the user gives none & Does not assume missing signal parameters; writing tools require them explicitly (e.g., \code{wrsamp -F 250 -G 200 -R 16 -U mV}); missing units are shown as \code{nd} (not defined) \\\hline
Finding files & Searches the WFDB path, ending with PhysioNet.org & Reads and writes files at specified locations: user-selected input and output directories, absolute or home (\code{\textasciitilde}) paths, or URLs \\\hline
Network access & Enabled in builds with \code{libcurl}; may fall back silently to PhysioNet.org & Disabled by default; when enabled, accesses only explicitly specified URLs \\\hline
Annotation sorting & Runs \code{sortann} through the shell; under some conditions, \code{sortann} silently drops annotations & Sorts within the library by time, annotation number and channel, without dropping annotations \\\hline
Output over input & Opening an output signal file can truncate a signal file being read & Refuses to overwrite an input signal file \\\hline
Write errors & Reports errors but exits with status 0 & Reports errors and exits with status 1 \\\hline
Crashes & Some malformed input causes crashes or machine-dependent results & Produces a defined result in every language (20 conformance cases) \\
\bottomrule
\end{tabularx}
\end{table}

\subsection{Signal parameters}
The C package assigns default sampling frequencies, gains, ADC resolutions and physical units when these are missing, and the WAV and AHA converters use their own fixed values. Such assumptions can silently affect subsequent analyses or records written from the data. \wfdbx{} instead requires these parameters and fails without them. For example, the header of record \code{100s} from the MIT-BIH Arrhythmia Database~\citep{Moody2005arrhythmia} does not specify signal units. The C command \code{rdsamp -p -v} labels its physical values \code{(mV)}, which is correct for this record but not a safe general assumption. \wfdbx{} displays \code{(nd)} (not defined), while legacy mode displays \code{(mV)}. To use the C defaults when writing a record, the user specifies them explicitly (or explicitly requested \code{--legacy} mode behavior):

\begin{lstlisting}[language=bash]
wrsamp -i data.txt -o rec -F 250 -G 200 -R 12 -U mV   # the C package's defaults, stated
\end{lstlisting}

\subsection{File locations and network access}
The C library searches the WFDB path and reads the first matching file, potentially from PhysioNet. It can also request optional \code{.info} files from the server, exposing local record paths. \wfdbx{} instead uses explicit file locations: absolute paths, paths under the user's home directory (\code{\textasciitilde}), URLs, or paths resolved against user-selected input and output directories. Relative signal-file names remain relative to their headers. The directories may point to local or network-accessible locations. Network access is supported in all seven languages but is disabled by default. For example, reading the first 10 seconds of MIT-BIH record 100 from PhysioNet requires:

\begin{lstlisting}[language=bash]
export WFDBX_INPUT_DIR=https://physionet.org/files/mitdb/1.0.0/
export WFDBX_NETWORK=1
rdsamp -r 100 -t 10
\end{lstlisting}

\subsection{Checksums, annotation sorting and write protection}
\wfdbx{} reports a checksum mismatch as a configurable warning rather than an error, allowing records with inconsistent checksums to be read. This is useful for records generated by software that computes checksums differently from the C package, including some versions of \code{wfdb-python} or WFDB for MATLAB and Octave. 

Annotation files are sorted within the library instead of invoking an external \code{sortann} process, and no annotations are discarded. An output signal file is also prevented from overwriting an input signal file (from the same run), regardless of how its path is specified. Write failures are reported with a nonzero exit status so that scripts can detect them.

\subsection{Legacy mode}
The original behavior remains available through \code{--legacy}, specified as the first argument of a command-line tool, or through the legacy option of the library and high-level functions. This preserves the original function and tool names without requiring separate legacy interfaces. Legacy mode reproduces the unmodified C package, including its default signal parameters, WFDB path handling and \code{sortann} behavior, and is tested with a separate conformance suite (Section~\ref{sec:verification}). It is intended for reproducing results obtained with the C tools with minimal changes to existing code. It is not recommended for new development because it retains the limitations and potentially unsafe behaviors that motivated \wfdbx{}.

\subsection{Additions}

\wfdbx{} also provides a high-level API, the \code{check\_record} function for checking WFDB compatibility, configurable warnings for checksum mismatches with results available through a record flag, and separately selectable input and output directories. The high-level API is recommended for new applications.

\section{AI-assisted codebase development}
\label{sec:development}
The initial \wfdbx{} codebase was developed interactively with Claude Code (model Claude Opus 5.5, Anthropic, in medium-effort mode), an AI coding assistant, under the author's direction. The author defined the requirements (Section~\ref{sec:goals}), architecture, order of work, and testing method; made all design decisions, including the departures from the C package (Section~\ref{sec:divergences}); reviewed the work at each stage; and independently validated the software on local and network machines (Linux and MacOS). The author also provided sample code from OSET~\citep{oset} and example codebases, which had identified design limitations and existing bugs in the existing MATLAB, wfdb-python, and C versions. The assistant wrote the code, tests, and documentation based exclusively on the WFDB C version 10.7.0, the public definitions of the WFDB file formats, and the additional codebases provided by the author and the author's instructions. After development, the code was reviewed by a second AI system, ChatGPT Work (model GPT 6.1 Sol, OpenAI, in medium-effort mode) for corner cases, and by human experts experienced with the WFDB file format and software package. Improvements were made iteratively to address implementation and performance issues. \wfdbx{} is maintained on Github.com, enabling further improvements and bug fixes by the user community.

\subsection{Instructions to the AI assistant}
Throughout the development phase, the author's messages to the assistant amounted to 250 prompts (messages) and 57,929 words, with a median length of 23 words; about 7,300 of these words were the author's own instructions, and the remaining 50,600 were pasted material, such as external reviews and terminal outputs.

To make the development process of \wfdbx{} reproducible, we summarize the main instructions given to the AI coding assistants. These instructions were designed to ensure that the implementations remained faithful to the original C sources. A discussion regarding the reliability of this methodology is presented in Section~\ref{sec:discussion}. The highlights of the instructions provided throughout the development were as follows:

\begin{itemize}
  \item \textit{Translate from the C code and WFDB format manual only; do not copy or reproduce other implementations:} The assistant was asked to translate the C sources of the WFDB Software Package 10.7.0 and the publicly available WFDB header and data format, and not to copy from the web or reproduce from memory the behavior of existing Python or MATLAB implementations of WFDB, since these differ from the C package (Section~\ref{sec:background}).
  \item \textit{The C version defines the expected results:} Expected test results were produced by running the C programs and the C library. The tests and their expected results were not to be changed to make an implementation pass; a suspected error in an expected result had to be reported, with evidence, rather than corrected automatically.
  \item \textit{Fix the cause, not the symptom and verify each design choice with the developer:} A correction had to be the smallest algorithmic or programming change that fixes the cause of an error, verified and applied in every language that had the error, and checked against the full test suites of these languages.
  \item \textit{Paraphrase author requests for confirmation and explain before changing:} The author requested the assistant to summarize and paraphrase each design requirement for author confirmation. Also, before any substantial change (a new feature, a departure from the C behavior or a change of design), the assistant had to explain the problem and its plan, and the author decided whether and how to proceed.
  \item \textit{Code and documentation standards:} No dependency beyond the standard library of each language (and NumPy in Python); no unexplained numerical constants; in every source file, a header naming the C file translated, its original authors, license and the changes from the C code; documentation in plain human-readable language, consistent across the seven languages.
  \item \textit{No network access in tests:} Tests could contact no host other than the local machine, so that their results do not depend on the network or model memory.
\end{itemize}

\subsection{Order of work}

The implementations were developed one language at a time. Each had to pass the full conformance suite before the next was started.

\begin{enumerate}
  \item \textit{Test infrastructure and Python:} A reference build of the C package (with FLAC support), the language-independent conformance harness and the test records were set up first. Python was then translated tool by tool, starting with \code{rdsamp}; the conformance cases of each tool were written and passed before the next tool was started. The Python implementation became the reference; its structure and conventions were used as templates for the other languages.
  \item \textit{C++, MATLAB/Octave, Rust, Julia, and R} were implemented in this order. Each was translated from the C sources, following the structure of the Python implementation, and checked against the same conformance cases. Unit tests of every library function, with expected values printed by the C library, were then added in each language.
  \item \textit{Kotlin Multiplatform} was added later for mobile and JVM applications, in four steps (foundation, signals and FLAC, tools, high-level API), and evaluated by the same tests.
  \item \textit{Review of all languages:} Every function was checked against its C original. Apparent errors in the C code were checked against the WFDB documentation, to distinguish errors from intended behavior. The code was checked for quality (e.g., for unexplained or hard-coded constants).
  \item \textit{Deliberate departures from the C package:} The changes of Section~\ref{sec:divergences} (no built-in signal parameters, files only where they are named, network access off by default, sorting without loss, protection of input files, write errors in the exit status) and legacy mode were made one at a time. Each was first implemented in the Python implementation and in the patched C reference that produces the expected results, and then in the other languages.
  \item \textit{Code optimization:} Once all implementations passed the tests, the assistant was asked to make each one as fast as its language allows, without changing any result. The changes followed the best coding practices of each language: reading signal files through a block reader rather than one frame per call; processing samples in small blocks that stay in the processor's cache (C++, Rust, Julia and Kotlin) or with whole-array operations instead of loops over indices (Python, R and MATLAB/Octave); allocating each result array once, from the length given in the header, rather than growing it; converting to physical units without temporary arrays; writing whole blocks of frames at once rather than one frame per call (Julia, R and MATLAB/Octave); formatting the output of \code{rdsamp} a block at a time; and, for the Kotlin tools, a class-data-sharing archive that halves the start-up time of the JVM. Each optimized version was checked against the code it replaces by randomized equivalence tests, and all implementations were benchmarked against each other, against \code{wfdb-python} and against the original C library (Section~\ref{sec:performance}).
  \item \textit{External reviews and corrections:} Independent reviewers, a second AI assistant (GPT-6.1). Each reported problem was reproduced, classified, and corrected in every language that had it, with a new conformance case or unit test. In the last rounds, a \textit{differential fuzzer} (Section~\ref{sec:diff_fuzzer}) compared the implementations with the C programs on random records and tool options.
  \item \textit{Documentation and release:} The documentation of each language was rewritten in a common structure, with examples and example notebooks. Every source file was attributed to the authors of the C file it translates; \wfdbx{} itself is distributed under the GNU General Public License, version 3 or later. the Python package was published on PyPI.
\end{enumerate}

\noindent From the departures of item~4 onward, each round of changes was made first in the Python implementation, and then carried to the other languages in parallel: a separate instance of the assistant worked on each language from a common written specification, could modify only the directory of its own language, and had to pass the conformance suite, the legacy suite, the unit tests of its language and, in the later rounds, the differential fuzzer (Section~\ref{sec:verification}) before its work was accepted. In the last round of corrections, for example, each language had to show no difference from the C programs in at least 2,000 random cases (and 1,000 in legacy mode) for C++, Rust and Python, and 600 (and 300) for the other languages.

\subsection{Review and validation}

In addition to automated verification, \wfdbx{} was reviewed by independent human code reviewers and tested in several rounds by early users with backgrounds in biomedical signal processing and computer science. These reviewers had prior experience with the WFDB C package, wfdb-python, or the WFDB Toolbox for MATLAB and Octave, and tested public releases on their own records and applications. They included researchers working with PhysioNet data, organizers of and participants in PhysioNet Challenges, and developers adapting ICU bedside-monitor converters and recording viewers to the Rust, C++, and Python implementations. Their reports identified minor defects, slow operations on multi-segment records, unclear documentation, and design issues. Several design decisions described in this paper and the optimized reading paths of Section~\ref{sec:architecture} resulted from this feedback.

The author reviewed every report before making changes. Each issue was reproduced and classified as a defect in \wfdbx{}, behavior inherited from the C package, or intended behavior that was not sufficiently documented. Confirmed defects were corrected in all affected language implementations, with conformance cases or unit tests added to prevent recurrence. The author also reviewed proposed design changes and decided whether to retain or revise the design or preserve the original C behavior as a legacy option. All reported issues were addressed.

The author independently validated the software by installing the Python package from both the PyPI test index and PyPI and running its tests on separate macOS and Linux systems, running the documented examples, comparing results with those from users' applications, and reviewing the documentation for each implementation.

\subsection{Codebase size}
Table~\ref{tab:loc} provides the size of each implementation in the current \wfdbx{} version 1.1.0. For comparison, the part of the C package that \wfdbx{} translates (the library and the 33 programs) has about 22,300 lines. The implementations are larger, mainly because each also contains its own FLAC decoder and encoder (which the C package takes from libFLAC), the layer reproducing C behavior (Section~\ref{sec:architecture}), the high-level API, \code{check\_record}, and documentation comments. In every language, about 15\% of the codebase belongs to the tests.

\begin{table}[tb]
\centering
\small
\caption{Lines of code of each implementation (comments and blank lines included), counted in the source files of the \wfdbx{} repository, version 1.1.0 (October 2026).}
\label{tab:loc}
\begin{tabular}{lrrr}
\toprule
Implementation & Library and tools & Tests & Total \\
\midrule
Python        & 22,992 & 5,147 & 28,139 \\
C++           & 27,705 & 6,519 & 34,224 \\
Rust          & 37,444 & 6,966 & 44,410 \\
Julia         & 29,058 & 5,408 & 34,466 \\
R             & 23,562 & 5,012 & 28,574 \\
MATLAB/Octave & 41,358 & 6,446 & 47,804 \\
Kotlin        & 29,744 & 6,891 & 36,635 \\
\midrule
Total         & 211,863 & 42,389 & 254,252 \\
\bottomrule
\end{tabular}
\end{table}

\section{Verification}
\label{sec:verification}

\begin{figure}[tb]
  \hspace*{-20mm}%
  \begin{minipage}{\dimexpr\textwidth+20mm\relax}
    \includegraphics[trim={0 2cm 0 0}, clip,width=\linewidth]{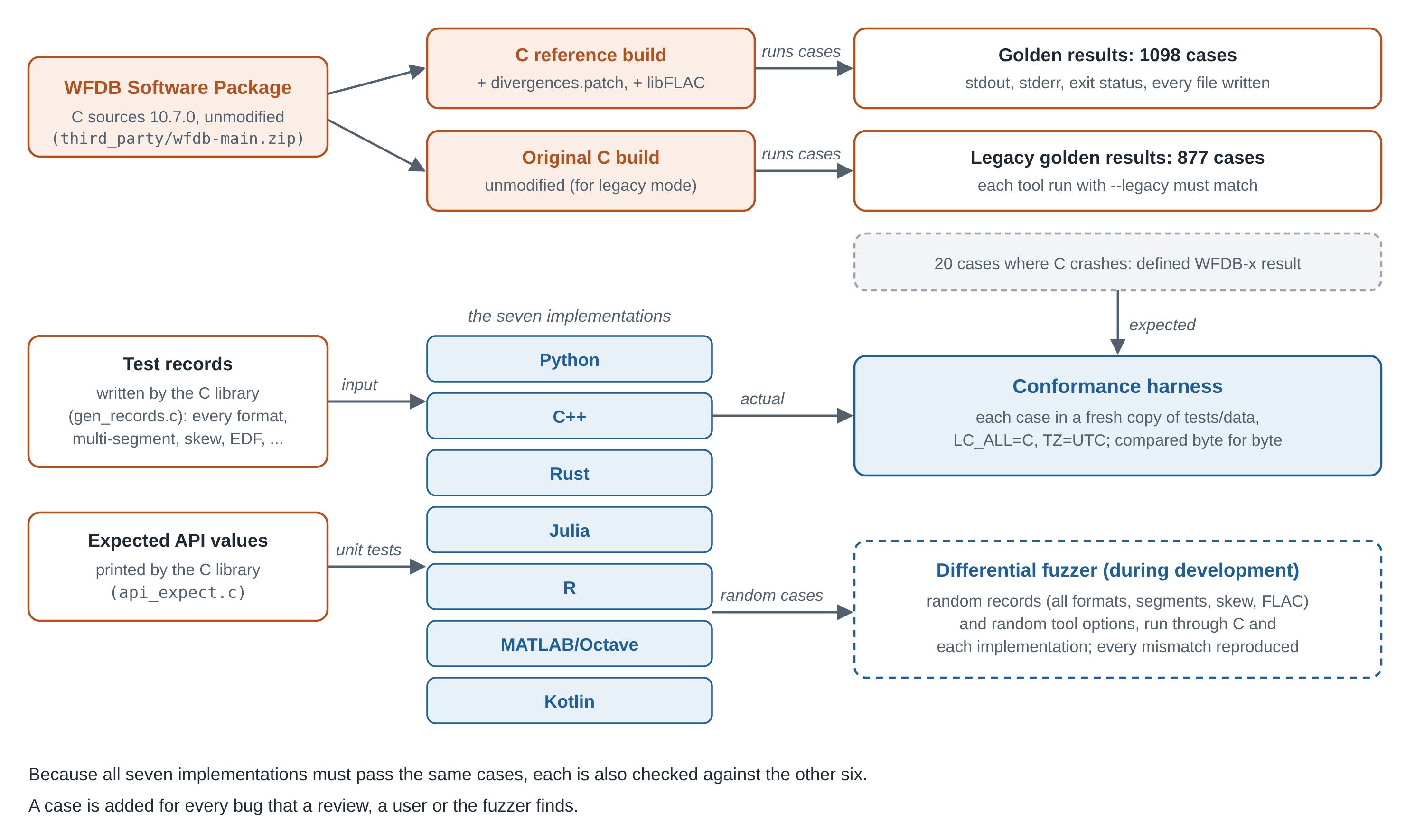}
    \caption{How \wfdbx{} is verified. The expected results of the conformance suite are produced by the C programs: a build with the deliberate departures applied, for the normal suite, and the unmodified package, for the legacy suite. Each implementation must reproduce them byte for byte.}
    \label{fig:verification}
  \end{minipage}
\end{figure}

\subsection{The conformance suite}
The main test of \wfdbx{} is a language-independent conformance suite containing 1098 test cases (Figure~\ref{fig:verification}). Each case runs a command-line tool with fixed arguments, input data, and environment settings, using a fresh copy of the test data. The test program then compares the results against the expected outputs, byte for byte, including standard output, error messages, exit status, and every file created or modified by the tool. All cases use the same locale and time-zone settings (\code{LC\_ALL=C}, \code{TZ=UTC}) to ensure reproducibility. Write errors are tested by directing output to \code{/dev/full}, a special Linux device file that always acts as if a storage device is completely full and rejects every write, simulating conditions such as a full disk. These tests run on Linux and are skipped on systems where \code{/dev/full} is unavailable.

The expected results were produced by running the C programs themselves: the unmodified sources of version 10.7.0, built with FLAC support and with the patch that applies the changes of Section~\ref{sec:divergences}. The only exceptions are 20 cases in which the C programs crash or give results that depend on the machine. For these, \wfdbx{} defines the correct result, which was recorded from one implementation and is used across all seven languages.

A second suite tests legacy mode. It runs 877 cases with \code{--legacy} flag as the first argument of each tool, and compares the results with those of the unmodified C package, so that legacy mode must reproduce the original package exactly.

All seven implementations pass all 1098 normal cases and all 877 legacy cases on the reference platform: x86-64 Linux with glibc, Python 3.13, GCC 13, Rust 1.97, Julia 1.13, R 4.3, GNU Octave 8.4 and OpenJDK 21. Because all seven are checked against the same expected results, the suites also check that the seven implementations agree with each other.

Any bugs reported or failing test cases identified in the future will be addressed with a similar methodology across all languages, and the corresponding test cases will be added to the test suite.

\subsection{Test records}

The test records were written by the C library itself, using a small C program, so that their contents are exactly what the C package produces. Together, they cover every signal storage format, records with signals at different sampling rates, records made of several segments (including empty and missing segments), signals that are shifted in time relative to each other (skew), signal files with a header or a fixed block size, EDF and EDF+ files, and annotation files that use every feature of the annotation format. Additional input files in other formats (WAV, AHA, SCP-ECG, and headers in older formats) test the format converters.

\subsection{Unit tests}

The conformance suite tests the command-line tools. The library functions and the high-level API are tested separately, by the unit tests of each implementation. The expected values of the library tests were printed by the C library, not written by hand. The tests of the high-level API check that a record written and read back is unchanged, that invalid arguments are rejected before any file is created (for example, a sampling frequency that is not a positive number), and that the library is left in a clean state after an error. In total, the unit tests contain about 9,350 assertions, each comparing one result with its expected value: between about 1,100 and 1,600 in each language (Table~\ref{tab:unittests}). Since many assertions run inside loops over records, formats or parameter values, the number of checks performed when the tests run is much larger.

\begin{table}[tb]
\centering
\small
\caption{Unit tests of each implementation: number of test files and of assertion statements in the test source code (counted in the Github repository as of \wfdbx{} version 1.1.0; assertions inside loops are counted once).}
\label{tab:unittests}
\begin{tabular}{lrr}
\toprule
Implementation & Test files & Assertions \\
\midrule
Python        & 10 & 1,088 \\
C++           & 8  & 1,181 \\
Rust          & 8  & 1,439 \\
Julia         & 6  & 1,495 \\
R             & 16 & 1,231 \\
MATLAB/Octave & 19 & 1,332 \\
Kotlin        & 19 & 1,587 \\
\midrule
Total         & 86 & 9,353 \\
\bottomrule
\end{tabular}
\end{table}

\subsection{Differential fuzzing}
\label{sec:diff_fuzzer}
In the later stages of development, a differential fuzzer\footnote{Differential fuzzing is a software testing technique that feeds the same generated inputs into two or more different implementations of the same functions that should behave identically and compares their outputs.} compared the implementations with the C programs on inputs that no one had written by hand. It generated random records, covering every storage format, several samples per frame, time-shifted signals, and multi-segment records with empty or unreadable segments. It then ran a randomly chosen tool with random options on each record, both with the C programs and with each implementation, and compared the results in the same way as the conformance suite. Every difference was reproduced and examined. When only some implementations differed from C, the cause was an error in \wfdbx{}, which was corrected. When all seven agreed with each other but differed from C, the cause was in the C code: these cases revealed reads of uninitialized or freed memory and integer overflows in the C programs, for which \wfdbx{} now defines and documents a result. Each difference found was added to the conformance suite as a new case. The differential fuzzer itself is not part of the repository; the cases it produced are.

\subsection{Documentation of behaviors}
The repository includes three documents that describe the behavior of \wfdbx{} and how it was tested: 1) \code{DIVERGENCES.md} lists the deliberate departures from the C package (Section~\ref{sec:divergences}) and the reason for each; 2) \code{KNOWN\_C\_BEHAVIORS.md} documents three kinds of C behavior: behaviors that \wfdbx{} reproduces although they are not described in the WFDB documentation or differ from it; behaviors that depend on the platform, such as the operating system or the C library; and undefined behaviors of the C code (for example, reads of uninitialized memory), for which \wfdbx{} defines and documents a result; 3) \code{INVENTORY.md} lists every command-line tool and library function, with its test status in each language.
\section{Performance}
\label{sec:performance}
We benchmarked the execution time and memory use of the seven implementations, the original C library and command-line tools, and wfdb-python 4.3.1 on a single machine: an x86-64 Linux virtual machine with 4 cores of an Intel Xeon at 2.1\,GHz. The C library and tools were the unmodified WFDB Software Package 10.7.0, built by its own \code{configure} script with its own compiler options (\code{-g -O}), as a user would install it. The benchmarks used three synthetic records modeled on real datasets, containing 12.5 million samples in total: \code{long}, two ECG signals in format 212 at 360\,Hz for one hour (1,296,000 frames), like the MIT-BIH records, with an annotation file of 4,500 beat annotations; \code{mimic}, a multi-segment record of 12 segments and 4 signals at 125\,Hz, the first at 500\,Hz, modeled on records from the MIMIC database; and \code{wide}, a 12-lead ECG in format 16 at 500\,Hz for 10 minutes.

\begin{figure}[t]
  \centering
  \includegraphics[width=\textwidth]{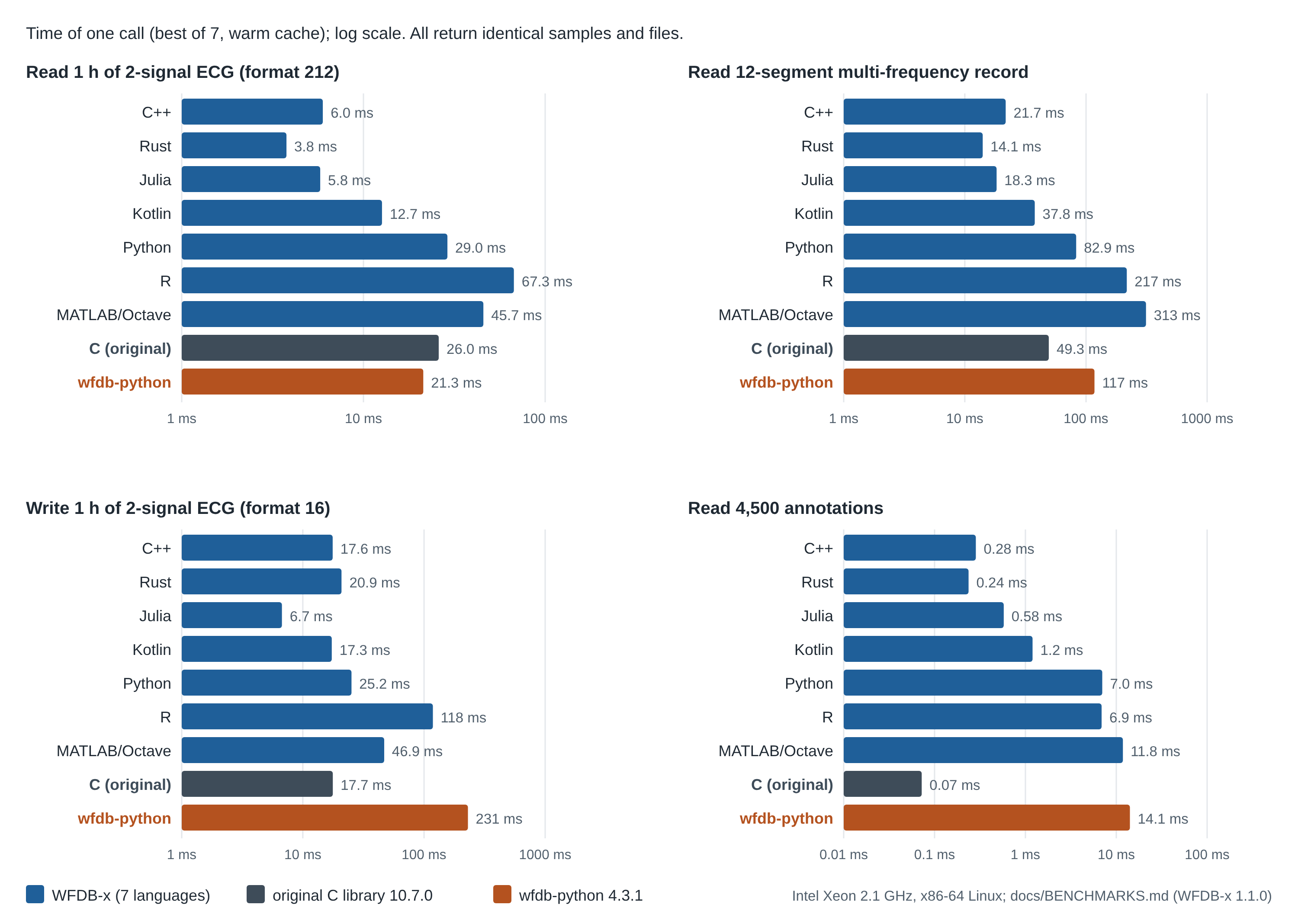}
  \caption{Time of one call in each \wfdbx{} implementation (high-level API), in the original C library (its library functions) and in wfdb-python 4.3.1, for four operations (best of 7 calls, and the better of two complete runs; files in the operating system's cache) in \textbf{logarithmic scale}. The values are those of Table~\ref{tab:bench}. All returned identical samples and annotations and wrote identical files.}
  \label{fig:benchmarks}
\end{figure}

Each \wfdbx{} implementation performed each operation with one call of its high-level API (\code{read\_record}, \code{write\_record}, \code{read\_annotations}, \code{write\_annotations}). The C library has no high-level functions, so a short C program performed the same operations with the library functions that a C program uses: \code{isigopen} and \code{getvec}, one frame per call, to read; \code{aduphys} for physical units; \code{osigfopen}, \code{putvec} and \code{newheader} to write; and \code{annopen}, \code{getann} and \code{putann} for annotations. To compare the libraries themselves, \code{long} was also read frame by frame with \code{getvec} through \wfdbx{}'s C++ library functions, with the same code as in C. Each library time is the fastest of 7 calls after one untimed call, taking the better of two complete runs. Every program saved the result of each operation, and the benchmark program stopped if any two implementations, the C library included, disagreed. All results were identical, value for value and byte for byte, in every run. Figure~\ref{fig:benchmarks} and Tables~\ref{tab:bench} and~\ref{tab:vsc} summarize the results.

\begin{table}[t]
\centering
\small
\caption{Time of one call (ms) and peak memory of the process (MB). C is the original WFDB C library, called through its library functions. The samples of \code{long} take 10\,MB.}
\label{tab:bench}
\setlength{\tabcolsep}{3.5pt}
\resizebox{\textwidth}{!}{%
\begin{tabular}{lrrrrrrrrr}
\toprule
 & C++ & Rust & Julia & Kotlin & Python & R & MATLAB/ & C & wfdb- \\
 &  &  &  &  &  &  & Octave & (original) & python \\
\midrule
Read \code{long} (ADC units) & 6.0 & 3.8 & 5.8 & 12.7 & 29.0 & 67.3 & 45.7 & 26.0 & 21.3 \\
Read \code{long} (physical units) & 12.9 & 14.8 & 12.1 & 17.1 & 37.2 & 99.5 & 71.9 & 30.1 & 36.4 \\
Read \code{mimic} & 21.7 & 14.1 & 18.3 & 37.8 & 82.9 & 217 & 313 & 49.3 & 117 \\
Read \code{wide} & 9.8 & 6.3 & 13.2 & 11.5 & 7.9 & 92.3 & 55.5 & 19.4 & 15.1 \\
Write \code{long} (format 16) & 17.6 & 20.9 & 6.7 & 17.3 & 25.2 & 118 & 46.9 & 17.7 & 231 \\
Read 4,500 annotations & 0.28 & 0.24 & 0.58 & 1.2 & 7.0 & 6.9 & 11.8 & 0.07 & 14.1 \\
Write 4,500 annotations & 0.14 & 0.17 & 0.34 & 1.1 & 0.86 & 3.8 & 6.3 & 0.14 & 6.9 \\
Read \code{long} frame by frame (\code{getvec}) & 23.9 & & & & & & & 21.6 & \\
\midrule
First call: read \code{long} & 14.4 & 9.4 & 57.3 & 75.8 & 142 & 102 & 95.8 & 32.7 & 123 \\
\midrule
Memory: read \code{long} (MB) & 24 & 13 & 311 & 149 & 80 & 157 & 115 & 12 & 118 \\
Memory: read \code{mimic} (MB) & 35 & 17 & 313 & 168 & 69 & 221 & 165 & 17 & 159 \\
\midrule
Tool: \code{rdsamp -r long} & 40.1 & 51.4 & 374 & 179 & 395 & 750 & 570 & 210 & \\
Tool: \code{rdann -r long -a atr} & 3.4 & 3.7 & 302 & 102 & 58.3 & 240 & 127 & 2.5 & \\
Tool: \code{wfdbdesc long} & 1.8 & 1.9 & 337 & 60.7 & 32.8 & 230 & 87.0 & 1.3 & \\
\bottomrule
\end{tabular}}
\end{table}

\begin{table}[t]
\centering
\small
\caption{Speed relative to the original C library: the time of the C library divided by the time of each \wfdbx{} implementation (from the unrounded times of Table~\ref{tab:bench}). Above 1, \wfdbx{} is faster; below 1, the C library is faster.}
\label{tab:vsc}
\setlength{\tabcolsep}{5pt}
\begin{tabular}{lrrrrrrr}
\toprule
 & C++ & Rust & Julia & Kotlin & Python & R & MATLAB/Octave \\
\midrule
Read \code{long} & 4.3 & 6.9 & 4.5 & 2.1 & 0.90 & 0.39 & 0.57 \\
Read \code{long} (physical units) & 2.3 & 2.0 & 2.5 & 1.8 & 0.81 & 0.30 & 0.42 \\
Read \code{mimic} & 2.3 & 3.5 & 2.7 & 1.3 & 0.59 & 0.23 & 0.16 \\
Read \code{wide} & 2.0 & 3.1 & 1.5 & 1.7 & 2.4 & 0.21 & 0.35 \\
Write \code{long} & 1.0 & 0.85 & 2.6 & 1.0 & 0.70 & 0.15 & 0.38 \\
Read 4,500 annotations & 0.25 & 0.31 & 0.13 & 0.06 & 0.01 & 0.01 & 0.01 \\
Write 4,500 annotations & 1.0 & 0.82 & 0.41 & 0.13 & 0.16 & 0.04 & 0.02 \\
\bottomrule
\end{tabular}
\end{table}

\subsection{Comparison with the original C library:} C++, Rust and Julia read records 1.5 to 6.9 times faster than the C library: the hour of ECG takes 3.8 to 6.0\,ms, against 26.0\,ms in C. Kotlin reads records 1.3 to 2.1 times faster than C. The Python implementation reads the 12-lead record 2.4 times faster than C and the other records 1.1 to 1.7 times slower, while R and MATLAB/Octave read them 1.8 to 6.3 times slower. The difference in reading speed comes from \wfdbx{}'s block reader (Section~\ref{sec:architecture}), not from the language: read frame by frame with \code{getvec}, \wfdbx{}'s C++ library and the C library take the same time (23.9 and 21.6\,ms, within the variation between runs). Writing a record takes the same time in C, C++ and Kotlin, and Julia is 2.6 times faster. The C library reads annotation files fastest, in 0.07\,ms for 4,500 annotations against 0.24 to 0.58\,ms in Rust, C++ and Julia: \code{getann} returns each annotation as it is decoded, whereas \code{read\_annotations} also builds the mnemonic (annotation symbol) of each annotation and the arrays of its result. C, C++ and Rust write annotation files in the same time (0.14 to 0.17\,ms). A build of the C library with \code{-O2} gave the same times, within the variation between runs (the hour of ECG in 24.5\,ms).

\subsection{Reading and writing across languages:} Rust, Julia and C++ are the fastest implementations. Kotlin reads and writes records 1.0 to 2.1 times slower than C++, once the JVM has compiled the code, and annotation files 4 to 8 times slower. Compared with wfdb-python, the Python implementation is 1.4 times slower on the single-segment record in ADC units and equally fast in physical units, and it is faster on every other operation: 1.4 times on the multi-segment record, 1.9 times on the 12-lead record, 2.0 times for reading annotations, 9.2 times for writing records and 8.0 times for writing annotations. R and MATLAB/Octave read the hour-long ECG in 46 to 67\,ms.

\subsection{Start-up and memory:} The first calls in Julia and Kotlin involve compilation and take 2 to 30 times longer than subsequent calls. This overhead has less impact in long-running applications. The C library, Rust and C++ use the least memory, 12 to 35\,MB for a record whose samples take 10\,MB; Julia's 310\,MB is mostly its runtime and compiler.

\subsection{Command-line tools:} The C++ and Rust \code{rdsamp} print the hour of ECG 4.1 to 5.3 times faster than the C \code{rdsamp} (40.1 and 51.4\,ms against 210\,ms), and the Kotlin \code{rdsamp} 1.2 times faster. For small tasks, start-up time dominates: the C tools start fastest (1.3\,ms for \code{wfdbdesc}), followed by the C++ and Rust tools (1.8 and 1.9\,ms), while the tools of the other languages pay the start-up time of their runtime (from 32.8\,ms in Python to 337\,ms in Julia).

\subsection{Limitations of the measurements:} All results come from one shared virtual machine, whose times vary by up to about 30\% between sessions; the ratios between implementations are more reliable than the absolute times. The processes were not pinned to a CPU core, and the tables give best times, not medians with confidence intervals. The files were in the operating system's cache. Peak memory includes about 11\,MB inherited from the benchmark runner, so values near it (C and Rust) are upper bounds. The MATLAB/Octave code ran in GNU Octave only because MATLAB license was unavailable on the tested VM, and the Kotlin code on the JVM only. The benchmark programs, including the C driver, are shared in the \wfdbx{} repository, so that the measurements can be repeated on other machines.

\section{Token and energy utilization of the AI-assisted development} 
The AI-assisted development of the core codebase of \wfdbx{} version 1.0.0 (the initial release) took 15 days, and consisted of one main Claude Opus 5.5 assistant session and 166 sub-agent sessions (excluding prior manual coding, bug detection, experimental setups, and architectural design by the author and team). The main session interacted with the author, wrote specifications for the sub-agents, assigned each sub-agent a single task (for example, translating one language or completing one round of corrections in one language), and reviewed and integrated the results. Together, the sessions made 18,655 model calls and generated 10.0 million output tokens, including code, tests, documentation, commands, and the model's intermediate reasoning. They read 4.73 billion input tokens because each call reads the full context of its session again, averaging about 254,000 tokens per call. Of these input tokens, 97.5\% (4.61 billion) were read from the prompt cache, 2.5\% (117 million) were written to it (104.6 million to the five-minute cache and 12.2 million to the one-hour cache), and fewer than 63,000 were read without caching. Table~\ref{tab:tokens} breaks these figures down by language according to each sub-agent's task. The main session accounts for a quarter of the output and most of the Python implementation, which the main agent wrote before sub-agents were used for the other languages; the sub-agents listed in the Python row worked only on later changes. Kotlin, the last language added, has the largest output per sub-agent (about 100,000 tokens, compared with 30,000 to 50,000 for the other languages) because it was written in four large tasks, each covering a major part of the implementation: the foundation, signals and FLAC, tools, and the high-level API. The row ``Several languages'' groups tasks that were not limited to one language: extensions to the C reference and conformance suite, reference implementations of coordinated changes, cross-language corrections, documentation reviews, and benchmarks. The final codebase contains about 254,000 lines across the seven languages (Table~\ref{tab:loc}), corresponding to about 40 output tokens per final line. This ratio reflects the tests, review rounds, and code rewritten during development.

At Anthropic's list prices for Claude Opus 5.5~\citep{anthropic2026pricing,anthropic2026caching} (\$4 per million uncached input tokens, \$20 per million output tokens, \$0.20 per million tokens read from the cache, and \$5 or \$8 per million tokens written to the five-minute or one-hour cache), these tokens correspond to an API-equivalent cost of about \$1,740: \$920 for reading cached context, \$620 for writing to the cache, and \$200 for output. Although input tokens outnumber output tokens by a factor of almost 500, they cost only about eight times as much as the output, because nearly all of them were cache reads, billed at one twentieth of the uncached input price.

Anthropic does not publish the energy consumed per token by its models, so the energy of the development can only be estimated. We used the per-token estimates of \citet{couch2026}, who derived them from Epoch AI's estimate of the energy of a typical language-model query and from the ratios between Anthropic's prices for the different kinds of tokens: about 390\,Wh per million uncached input tokens, 39\,Wh per million tokens read from the cache, and 1,950\,Wh per million output tokens. For tokens written to the cache, which these estimates do not list, we applied the same price ratios (1.25 times the uncached input price for the five-minute cache and twice for the one-hour cache), that is, 487.5 and 780\,Wh per million tokens. With these values, the development consumed an estimated 260\,kWh (Table~\ref{tab:tokens}): 180\,kWh for reading cached context, 60\,kWh for writing to the cache, and 20\,kWh for generating output\footnote{For comparison, according to the U.S. Energy Information Administration's 2020--2024 Residential Energy Consumption Survey, a four-person household in the United States consumes an average of 13,072\,kWh of electricity annually, equivalent to approximately 1,089\,kWh per month~\citep{EIA2024RECS}. Our AI-assisted coding token utilization totaled 24\% of this monthly utilization.}. Reading the context dominates, because each of the 18,655 calls read on average about 254,000 tokens. This estimate is uncertain by a large factor: the per-token values were not derived for Claude Opus 5.5; they assume that energy is proportional to price (Claude Opus 5.5 prices cache reads at 0.05 times the input price, not the 0.1 assumed in the estimate); and they exclude the energy of the computers that ran the builds and tests, as well as data-center overheads beyond those included in the original estimate. At the average electricity price in the United States as of 1 November 2026 (about 0.183 USD/kWh), 260\,kWh corresponds to about USD~48.

Token usage for the second AI assistant (ChatGPT GPT-6.1, the code reviewer) was approximately 46.3 million tokens, comprising 45.1 million cached input tokens, 1.03 million uncached input tokens, and 0.12 million output tokens, including reasoning. Subsequent adjustments, updates, and reviews are not included in the above figures.

Meanwhile, in total, the author spent approximately USD~160 on Claude cloud and ChatGPT credits for this development and review, which was above the estimated energy cost but significantly below the token-based cost estimate, suggesting that the actual power consumption per Claude token is significantly lower than publicly reported estimates.

\begin{table}[tb]
\centering
\small
\caption{Token utilization, API-equivalent cost and estimated energy of the AI-assisted development, from 15 to 29 October 2026, by session and by the language on which each sub-agent worked. Input tokens include the context read from and written to the prompt cache. Cost is computed at Anthropic's list prices for Claude Opus 5.5; energy is estimated from third-party per-token values~\citep{couch2026} (see text) and is uncertain by a large factor.}
\label{tab:tokens}
\setlength{\tabcolsep}{4pt}
\begin{tabular}{lrrrrrrr}
\toprule
 & Sub- & Model & \multicolumn{2}{c}{Output tokens} & Input tokens & Cost & Energy estimate \\
 & agents & calls & (millions) & (\%) & (millions) & (USD) & (kWh) \\
\midrule
Main session      & --  & 3,205  & 2.54  & 25.4 & 1,385 & 439   & 70 \\
Kotlin            & 13  & 1,143  & 1.30  & 12.9 & 264   & 99    & 15 \\
MATLAB/Octave     & 26  & 2,776  & 1.18  & 11.8 & 625   & 240   & 35 \\
R                 & 24  & 2,589  & 1.04  & 10.4 & 608   & 237   & 35 \\
Julia             & 25  & 2,090  & 0.85  & 8.5  & 406   & 162   & 23 \\
C++               & 21  & 1,549  & 0.78  & 7.8  & 320   & 126   & 18 \\
Rust              & 22  & 1,917  & 0.78  & 7.8  & 396   & 171   & 24 \\
Python            & 13  & 1,246  & 0.62  & 6.2  & 244   & 85    & 13 \\
Several languages & 22  & 2,140  & 0.91  & 9.1  & 482   & 184   & 27 \\
\midrule
Total             & 166 & 18,655 & 10.01 & 100  & 4,729 & 1,743 & 260 \\
\bottomrule
\end{tabular}
\end{table}

\section{Discussion}
\label{sec:discussion}

\subsection{The need for a cross-language WFDB implementation with identical behavior}

The widespread use of WFDB in PhysioNet databases, public data challenges, published algorithms and clinical research makes consistent reading and writing of these records important. Sharing the same file format is not sufficient if different software returns different samples, annotations, timing or physical values. This problem is particularly relevant when researchers develop algorithms in one language, validate them in another or transfer them to a deployed system. \wfdbx{} addresses this need by providing native implementations in seven languages, all tested against the original WFDB C code. The implementations also share a high-level API and the same 33 signal, annotation and conversion tools.

Compatibility does not mean retaining every historical behavior by default. Implicit sampling frequencies, gains and units may silently misrepresent a recording; file searches may retrieve an unintended copy or access the network; and annotation sorting may discard labels. \wfdbx{} addresses these problems while retaining the original C behavior in an explicitly selected legacy mode (Section~\ref{sec:divergences}). This is important for both new development and reproducibility: a researcher can use the revised behavior in a new application and still reproduce the behavior of older C-based workflows when necessary.

\subsection{Evidence for correctness and conformance}

The principal evidence for the consistency of \wfdbx{} is not that the seven implementations were generated from similar instructions, but that their results were tested against a common, independently executable reference. The original, human-written C library provides the expected behavior for the supported operations. All seven implementations pass 1098 normal conformance cases and 877 legacy cases, with comparisons of standard output, error messages, exit status and generated files. Separate unit tests contain 9,353 assertion statements across the languages, and randomized differential fuzzing compares the implementations with the C programs on records and options beyond the hand-written test cases (Section~\ref{sec:verification}). Optimized reading paths are also checked against their frame-by-frame counterparts. Human reviewers and early users have tested the public software on their own applications and recordings, and confirmed consistent behavior and improved speed.

These results are strong evidence of agreement for the tested inputs and operations, but they are not a formal proof of program equivalence. The seven codebases contain 254,252 lines in total, including tests, comments and blank lines (Table~\ref{tab:loc}); not every line has undergone individual human inspection. Nor can a finite collection of tests establish identical behavior for every valid or malformed input. Some C behaviors are undefined or platform-dependent, and \wfdbx{} explicitly defines a result for these cases rather than claiming equivalence to every possible C build. The scope and limitations of the conformance claim are therefore important: the implementations match the designated C reference on the tested cases, subject to the documented departures and the separate treatment of FLAC encoding.

The approach nevertheless provides a practical way to assess and maintain cross-language consistency. Because the original C sources, patched reference and test harness remain available, future human developers and AI coding tools can check changes against an established reference instead of relying on earlier AI-generated translations. A defect in one implementation can be reproduced, compared with C and corrected across the others. This does not prevent errors from propagating automatically, but it provides a means to detect and stop them before changes are accepted. The reference and tests are therefore at least as important to the long-term value of \wfdbx{} as the initial translations.

\subsection{AI-assisted development and performance}

The codebase was developed with AI assistance under the author's direction, including decisions on design, deviations from C, testing and acceptance of changes (Section~\ref{sec:development}). Large language models are stochastic, but the compilers and interpreters that build and run the codes are deterministic. Therefore, repeating the same prompts, even with the same settings, need not produce the same code. The documented instructions and development records make the process transparent, but they should not be interpreted as a recipe that will regenerate identical implementations. The more reproducible part of the methodology is the verification procedure, which can be rerun on the released source code without access to the original AI sessions.

The benchmarks also show that matching the C results need not impose a performance penalty. On the tested machine, C++, Rust and Julia read records 1.5 to 6.9 times faster than the original C library, mainly because the high-level API reads blocks instead of calling \code{getvec} for every frame. When C++ uses the same frame-by-frame interface as C, their reading times are comparable. The Python implementation performs competitively with \code{wfdb-python} and is substantially faster for writing records and annotations, although the original C library remains faster for reading annotations (Section~\ref{sec:performance}). These differences depend on the operation and execution environment. They were measured on one virtual machine with cached files and best-of-run timings, not across a representative range of computers or workloads.

The measured token use during code development, estimated cost and energy consumption describe the resources used in development, not the correctness of the result; the energy estimates are particularly uncertain.

The financial and environmental costs of AI-assisted development, including electricity consumption and associated carbon emissions, remain subjects of considerable debate. Based on our estimated token usage, the electricity consumed during the development of \wfdbx{} was approximately 24\% of the average monthly electricity consumption of a four-person household in the United States. However, this cost should also be considered against the potential long-term benefits of the software. WFDB and other physiological waveform formats are increasingly used in intensive care units, healthcare systems, and wearable monitoring devices, where large volumes of data are continuously recorded, transferred, and processed. Improvements in the efficiency of widely used data-processing and storage libraries can reduce computational demands, energy consumption, and cloud computing costs over repeated use. Even modest improvements per operation may become significant when applied to large datasets and millions of recordings. Although we have not yet quantified these downstream savings, the development of \wfdbx{} illustrates how AI-assisted software engineering may help improve computational tools whose cumulative resource consumption could substantially exceed the one-time cost of their development. Similar approaches may be useful for other widely deployed computational libraries and data-processing engines.

\subsection{Limitations and future validation}

The main limitation of \wfdbx{}'s current release is the breadth of validation. The full conformance suites were run on x86-64 Linux with glibc. Broader testing is needed on Windows and macOS, on different processor architectures and with different C libraries, particularly for operations with platform-dependent behavior. The MATLAB/Octave implementation has been tested in GNU Octave but still needs direct testing in MATLAB. Kotlin has been tested on the JVM, while its native iOS targets remain to be built and evaluated. Tests should also be expanded for unusual combinations of storage formats, multi-segment records, malformed files and failure conditions.

A further useful step would be systematic testing against many other WFDB datasets hosted on PhysioNet, rather than only the constructed and user-supplied records used here. Such testing could expose unsupported corner cases and identify records with inconsistent checksums, incorrect scaling metadata, or conversion errors introduced by different software implementations. These findings would need to distinguish an error in a dataset from an error in the reader or writer; \code{check\_record} and the original C tools provide starting points for that work. Extending the tests to new datasets, platforms and real applications is more meaningful than increasing the number of test cases without broadening their coverage.

Finally, \wfdbx{} is a recent project. It covers the WFDB library and 33 signal, annotation and conversion tools, not the C package's detectors, evaluation programs, spectral-analysis tools or WAVE viewer. Other tools, including OSET~\citep{oset}, address parts of that broader scope. Continued use, independent review and reports from the community will be essential. Future discrepancies should be reproduced against the reference, corrected in all affected languages and added to the tests. 

\section{Conclusion}

\wfdbx{} addresses a practical source of irreproducibility in physiological signal processing: different software implementations of the same WFDB format may read or write records differently. It provides native implementations in seven programming languages, with high-level APIs and command-line tools, using the original C code as a common reference. Agreement across the conformance suites, unit tests, differential fuzzing, and independent reviews supports the consistency of the implementations for the tested operations, without constituting a formal proof of equivalence. Documented changes remove several unsafe or ambiguous legacy behaviors, while an optional legacy mode preserves the original results when required. By keeping the reference code, test data, and verification procedures available, \wfdbx{} enables both human developers and future AI tools to check and improve the software while maintaining its connection to the established WFDB implementation developed by the late George B. Moody and his team.
\section*{Software availability}

\wfdbx{} is freely available as open-source software. Version 1.1.0, described in this paper, is available from \url{https://github.com/alphanumericslab/wfdb-x}. The repository contains the native implementations in seven programming languages, the command-line tools, installation instructions, examples, test data, unit tests, conformance suites, benchmark scripts, and documentation. The Python package is also distributed through the Python Package Index (PyPI) and can be downloaded from \url{https://pypi.org/project/wfdb-x/} or installed using \code{pip install wfdb-x}. The repository's \code{docs} directory documents the deliberate departures from the C package, reproduced C behaviors, verification procedures, benchmarks, and source-code attribution. Bug reports and contributions can be submitted through the GitHub repository.

\wfdbx{} is distributed under the GNU General Public License (GPL), version 3 or later. The unmodified WFDB C sources included in the repository for reference testing retain their original licenses: the GNU Library General Public License (LGPL), version 2 or later, for the library, and the GPL, version 2 or later, for the applications.

\section*{AI usage disclosure}

Generative AI was used in the development, testing, documentation, and review of \wfdbx{}. Claude Code (Claude Opus 5.5, Anthropic) was used for code development under the author's direction, and ChatGPT Work (GPT 6.1 Sol, OpenAI) was used for additional code review. The author defined the design and requirements, directed and reviewed the development, and independently validated the results. Human experts and early users also reviewed and tested the software. The development and verification procedures are detailed in Sections~\ref{sec:development} and~\ref{sec:verification}. The author reviewed and revised all AI-assisted work and takes full responsibility for the software and findings.

\section*{Acknowledgements}
\wfdbx{} would not exist without the WFDB Software Package, written by the late George B. Moody and maintained by Benjamin Moody, with contributions credited file by file in the \wfdbx{} sources, nor without PhysioNet and Ary L. Goldberger and the late Roger G. Mark. We thank the external reviewers and early users whose reports improved the software. The author acknowledges suppot from Georgia CTSA under the NIH-NCATS award UL1TR002378. The author declares no competing interests.

\bibliography{references}

\newpage
\appendix

\section{Reading and writing records and annotations}
\label{app:examples}

The examples use \code{100s}, the first minute of record 100 from the MIT-BIH Arrhythmia Database. It contains two ECG signals sampled at 360\,Hz (21,600 samples per signal) and 75 beat annotations. The record is included in the repository's \code{tests/data} directory and in the Python package. The examples write new records to the current directory. The values in the comments are those printed by the code.

\subsection{Python}

\begin{lstlisting}[language=Python]
import numpy as np
import wfdbx

# Read from the package's sample data; writes go to the current directory.
wfdbx.wfdbx_set_input_dir(wfdbx.sample_data_dir())

# Read waveforms: the whole record, in physical units.
rec = wfdbx.read_record("100s", physical=True)
print(rec.fs, rec.names, rec.signals.shape)   # 360.0 ['MLII', 'V5'] (21600, 2)

# One signal, from 10 s to 20 s (times as WFDB time strings or sample numbers).
seg = wfdbx.read_record("100s", channels=["MLII"], sampfrom="0:10", sampto="0:20")

# The header only, without the samples.
hdr = wfdbx.read_header("100s")
print(hdr.nsig, hdr.nsamples)                  # 2 21600

# Read annotations.
ann = wfdbx.read_annotations("100s", "atr")
print(len(ann), ann.symbols[:3], ann.samples[:3])   # 75 ['+', 'N', 'N'] [ 18  77 370]

# Write waveforms: 10 s of random data as record "rand" (rand.hea, rand.dat).
# The sampling frequency, gain, ADC resolution and units are required.
x = np.random.randn(3600, 2)
wfdbx.write_record("rand", x, fs=360, fmt=16, physical=True,
                   gains=1000, adcres=16, units="mV", names=["A", "B"])

# Write annotations for "rand" (rand.qrs).
wfdbx.write_annotations("rand", "qrs", samples=[100, 460, 820],
                        symbols=["N", "N", "V"])

# Read a record from PhysioNet: network access for this call only.
r100 = wfdbx.read_record("100", input_dir="https://physionet.org/files/mitdb/1.0.0/",
                         network=True, sampto="0:10")
\end{lstlisting}

\subsection{C++}

\begin{lstlisting}[language=C++]
#include <wfdbx/record.hpp>
#include <cstdio>

int main() {
    wfdbx::wfdbx_set_input_dir("tests/data");

    // Read waveforms: the first 10 seconds, in ADC units.
    wfdbx::ReadOptions opt;
    opt.to = "0:10";
    wfdbx::Record rec = wfdbx::read_record("100s", opt);
    std::printf("%g %d %zu\n", rec.fs, rec.nsig, rec.nsamples());   // 360 2 3600
    std::printf("%d %g\n", rec.at(0, 0), rec.physical(0, 0));       // 995 -0.145

    // Read annotations.
    wfdbx::Annotations ann = wfdbx::read_annotations("100s", "atr");
    std::printf("%zu %s %lld\n", ann.size(), ann.symbols[1].c_str(),
                ann.samples[1]);                                     // 75 N 77

    // Write waveforms: the 10 seconds as record "copy".
    wfdbx::WriteOptions w;
    w.gains = {200};
    w.baselines = {1024};
    w.adcres = {11};
    w.units = {"mV"};
    w.names = rec.names();
    wfdbx::write_record("copy", rec.samples, rec.nsig, rec.fs, w);

    // Write annotations for "copy".
    wfdbx::Annotations qrs;
    qrs.samples = {100, 460, 820};
    qrs.symbols = {"N", "N", "V"};
    wfdbx::write_annotations("copy", "qrs", qrs);
}
\end{lstlisting}

\subsection{Rust}

\begin{lstlisting}
use wfdbx::{read_annotations, read_record, write_annotations, write_record,
            Annotations, ReadOptions, WriteOptions};

fn main() -> Result<(), wfdbx::record::Error> {
    wfdbx::wfdbx_set_input_dir("tests/data");

    // Read waveforms.
    let rec = read_record("100s", &ReadOptions::default())?;   // ADC units
    println!("{} {} {}", rec.fs, rec.nsig, rec.nsamples());    // 360 2 21600
    let mlii = rec.physical_signal(0)?;                         // physical units
    println!("{:?}", &mlii[..3]);                               // [-0.145, -0.145, -0.145]

    // Read annotations.
    let ann = read_annotations("100s", "atr", None, None, None, false)?;
    println!("{} {:?}", ann.len(), &ann.samples[..3]);          // 75 [18, 77, 370]

    // Write waveforms: the first 10 seconds as record "copy".
    let opt = ReadOptions { to: Some("0:10".into()), ..Default::default() };
    let x = read_record("100s", &opt)?;
    let w = WriteOptions {
        gains: x.siginfo.iter().map(|s| s.gain).collect(),
        baselines: x.siginfo.iter().map(|s| s.baseline).collect(),
        adcres: vec![11],
        units: vec![Some("mV".into())],
        names: x.names(),
        ..Default::default()
    };
    write_record("copy", &x.samples, x.nsig, x.fs, &w)?;

    // Write annotations for "copy".
    let qrs = Annotations {
        samples: vec![100, 460, 820],
        symbols: vec!["N".into(), "N".into(), "V".into()],
        ..Default::default()
    };
    write_annotations("copy", "qrs", &qrs, 0.0, None)?;
    Ok(())
}
\end{lstlisting}

\subsection{Julia}

\begin{lstlisting}
using WFDBX
wfdbx_set_input_dir("/path/to/wfdb-x/tests/data")

# Read waveforms.
rec = read_record("100s")                          # ADC units
println(rec.fs, " ", size(rec.samples))            # 360.0 (21600, 2)
x = physical(rec)                                  # physical units
println(x[1, :])                                   # [-0.145, -0.065]

# Read annotations.
ann = read_annotations("100s", "atr")
println(length(ann), " ", ann.samples[1:3])        # 75 [18, 77, 370]

# Write waveforms: the first 10 seconds as record "copy".
r10 = read_record("100s"; to="0:10")
write_record("copy", r10.samples, r10.fs; fmt=16,
             gains=[s.gain for s in r10.siginfo],
             baselines=[s.baseline for s in r10.siginfo],
             adcres=11, units="mV", names=signal_names(r10))

# Write annotations for "copy".
write_annotations("copy", "qrs",
                  Annotations(samples=[100, 460, 820], symbols=[b"N", b"N", b"V"]))
\end{lstlisting}

\subsection{R}

\begin{lstlisting}[language=R]
library(wfdbx)
wfdbx_set_input_dir("wfdb-x/tests/data")

# Read waveforms.
rec <- read_record("100s", physical = TRUE)
rec$fs                     # 360
rec$names                  # "MLII" "V5"
dim(rec$signals)           # 21600 2

# Read annotations.
ann <- read_annotations("100s", "atr")
nrow(ann)                  # 75
head(ann$symbol, 3)        # "+" "N" "N"

# Write waveforms: the first 10 seconds as record "copy".
x <- read_record("100s", sampto = "0:10")     # ADC units
write_record("copy", x$signals, fs = x$fs, fmt = 16, gains = x$gains,
             baselines = x$baselines, adcres = 11, units = "mV", names = x$names)

# Write annotations for "copy".
write_annotations("copy", "qrs", sample = c(100, 460, 820),
                  symbol = c("N", "N", "V"))
\end{lstlisting}

\subsection{MATLAB and GNU Octave}

\begin{lstlisting}[language=Matlab]
wfdbx.wfdbx_set_input_dir('/path/to/wfdb-x/tests/data');

% Read waveforms.
rec = wfdbx.read_record('100s', 'physical', true);
rec.fs                      % 360
size(rec.signals)           % [21600 2]

% Read annotations.
ann = wfdbx.read_annotations('100s', 'atr');
numel(ann.samples)          % 75
ann.symbols(1:3)'           % {'+', 'N', 'N'}

% Write waveforms: the first 10 seconds as record 'copy'.
x = wfdbx.read_record('100s', 'sampto', '0:10');    % ADC units
wfdbx.write_record('copy', x.signals, x.fs, 'fmt', 16, 'gains', x.gains, ...
    'baselines', x.baselines, 'adcres', 11, 'units', 'mV', 'names', x.names);

% Write annotations for 'copy'.
wfdbx.write_annotations('copy', 'qrs', [100 460 820], 'symbols', {'N', 'N', 'V'});
\end{lstlisting}

\subsection{Kotlin}

\begin{lstlisting}
import wfdbx.*

fun main() {
    wfdbxSetInputDir("tests/data")

    // Read waveforms (ADC and physical units).
    val rec = readRecord("100s", physical = true)
    println("${rec.fs} ${rec.names} ${rec.nsamples}")    // 360.0 [MLII, V5] 21600

    // Read annotations.
    val ann = readAnnotations("100s", "atr")
    println("${ann.size} ${ann.symbols.take(3)}")        // 75 [+, N, N]

    // Write waveforms: the first 10 seconds as record "copy".
    val x = readRecord("100s", to = TimeSpec("0:10"))    // ADC units
    writeRecord("copy", x.samples, x.nsig, x.fs, fmt = 16,
                gains = x.gains.toDoubleArray(), baselines = x.baselines.toIntArray(),
                adcres = intArrayOf(11), units = listOf("mV"), names = x.names)

    // Write annotations for "copy".
    writeAnnotations("copy", "qrs", longArrayOf(100, 460, 820),
                     symbols = listOf("N", "N", "V"))
}
\end{lstlisting}

\subsection{The command-line tools}

The tools have the same names and options in every language. In C++ and Rust, they are installed as programs (\code{rdsamp}); in the other languages they run through a launcher (\code{wfdbx rdsamp} in Python, Julia, R and MATLAB/Octave, \code{kotlin/bin/wfdbx rdsamp} for Kotlin).

\begin{lstlisting}[language=bash]
# Read waveforms: the first 10 s, in physical units, with column headings.
rdsamp -r 100s -t 10 -p -v

# Read annotations: the first 30 s.
rdann -r 100s -a atr -t 0:30

# Write waveforms: a record "rec" from a text file of two columns.
wrsamp -i data.txt -o rec -F 360 -G 200 -R 11 -U mV 0 1

# Write annotations: copy the annotations of 100s to a new annotator.
rdann -r 100s -a atr | wrann -r 100s -a copy

# The same commands as the original package (built-in defaults, WFDB path).
rdsamp --legacy -r 100s -t 10 -p -v
\end{lstlisting}

\section{The command-line tools}
\label{app:tools}

\begin{table}[h]
\centering
\small
\begin{tabularx}{\textwidth}{>{\raggedright\arraybackslash}p{4.2cm}>{\raggedright\arraybackslash}X}
\toprule
Group & Tools \\
\midrule
Signals & \code{rdsamp}, \code{wrsamp}, \code{sigamp}, \code{snip}, \code{xform}, \code{wfdbcat} \\
Annotations & \code{rdann}, \code{wrann}, \code{sortann}, \code{ann2rr}, \code{rr2ann} \\
Records and time & \code{wfdbdesc}, \code{wfdbtime}, \code{time2sec}, \code{sampfreq}, \code{signame}, \code{signum}, \code{wfdbwhich} \\
Format converters & \code{edf2mit}, \code{mit2edf}, \code{rdedfann}, \code{wav2mit}, \code{mit2wav}, \code{wfdb2mat}, \code{a2m}, \code{m2a}, \code{ad2m}, \code{md2a}, \code{ahaecg2mit}, \code{parsescp}, \code{makeid}, \code{readid}, \code{revise} \\
\bottomrule
\end{tabularx}
\caption{The 33 command-line tools of \wfdbx{}: the 18 signal and annotation tools of the C package's \code{app} directory and the 15 converters of its \code{convert} directory, each available in all seven languages.}
\end{table}

\vfill

\end{document}